\documentclass[12pt]{article}

\begin{document}

\title{ Fast Codes for Large Alphabets.
 \footnote {Supported by the
INTAS under the Grant no. 00-738. } }

\author{ Boris Ryabko, Jaakko Astola, Karen Egiazarian.  }
\date{}
\maketitle

\begin{abstract}
We address the problem of constructing a fast lossless code in the
case when the source alphabet is large. The main idea of the new
scheme may be described as follows. We group letters with small
probabilities in subsets (acting as super letters) and use time
consuming coding for these subsets only, whereas letters in the
subsets have the same code length and therefore can be coded fast.
The
described scheme can be applied to sources with known and unknown
statistics.

\end{abstract}


\section{Introduction.}

The computational efficiency of lossless data compression  for
large alphabets has attracted attention of researches  for
ages  due to its great importance in practice. The alphabet
of $2^8 = 256$ symbols, which is commonly used in compressing
computer files, may already be treated as a large one, and with
adoption of the UNICODE the alphabet size will grow up to $2^{16}=
65536 $.
 Moreover, there are many data compression methods when
the coding is carried out in such a way that, first input data
are transformed by some algorithm, and then the resulting sequence
 is compressed by a lossless code. It turns out that
very often the alphabet of the sequence is very large or even
infinite. For instance, the run length code, many implementations
of Lempel- Ziv codes, Grammar - Based codes \cite{Ki1,Ki2} and
many methods of image compression can be described in this way.
That is why the problem of constructing high-speed codes for large
alphabets has attracted great attention by researches. Important
results have been obtained  by Moffat, Turpin
\cite{Moffat90,Moffat94,Moffat99,M-T1,M-T,T-M} and others
\cite{Jo,RyabkoDAN,Ryabko,Fenwick, R-Ri}.

For many adaptive lossless codes the speed of coding depends substantially
 on the alphabet size, because of the need to maintain
cumulative probabilities. The speed of an obvious (or naive)
method of updating the cumulative probabilities is proportional to
the alphabet size $N$. Jones \cite{Jo} and Ryabko \cite{RyabkoDAN}
have independently suggested two different algorithms, which
perform all the necessary transitions between individual and
cumulative probabilities in $O(\log N)$ operations under $ (\log N
+ \tau)$- bit words , where $\tau$ is a constant depending on
the redundancy required, $N$ is the alphabet size. Later many such
algorithms have been developed and investigated in numerous papers
\cite{Moffat90,Ryabko,Fenwick,Moffat94,Moffat99}.

In this paper we suggest a method for speeding up codes
 based on the following main idea. Letters of the alphabet are
put in order according to their probabilities (or frequencies of
occurrence), and the letters with  probabilities close to each others
are grouped
in subsets (as new super letters), which contain letters
with small probabilities. The key
point is the following: equal probability is ascribed
to all letters in one subset, and, consequently, their codewords
have the same length. This gives a possibility  to encode and
decode them much faster than if they are
different. Since each subset of the grouped letters
is treated as one letter in the new alphabet, whose size is much
smaller than the original alphabet.
Such a grouping can increase the redundancy of the code. It
turns out, however, that a large decrease in the alphabet size may cause a
relatively small increase in the redundancy. More exactly, we
suggest a method of grouping for which the number of the
groups as a function of the redundancy ($\delta$) increases as $c
( \log  N + 1/ \delta )+ c_1 $, where $N$ is the alphabet size and
$c, c_1$ are constants.

In order to explain the main idea we consider the following
example. Let a source generate letters $ \{ a_0,\ldots , a_4 \}$
with probabilities $ p(a_0) = 1/16,\, p(a_1) = 1/16, \,p(a_2) =
1/8, \,p(a_3) = 1/4, \,p(a_4) = 1/2, $ correspondingly. It is easy
to see that the following code $$ code(a_0) = 0000, code(a_1) =
0001, code(a_2) = 001, code(a_3) = 01, code(a_4) = 1 $$ has the
minimal average codeword length. It seems that for decoding one needs
to look at one bit for decoding $a_4$, two bits for
decoding $a_3$, 3 bits for $a_2$ and 4 bits for $a_1$ and $a_0$.
However,
consider another code $$ \widetilde{code}(a_4) = 1,
\widetilde{code}(a_0) = 000, \widetilde{code}(a_1) = 001,
\widetilde{code}(a_2) = 010, \widetilde{code}(a_3) = 011, $$ and we
 see that, on the one hand, its average codeword length is a
little larger than in the first code (2 bits instead of 1.825
bits), but, on the other hand, the decoding is simpler. In fact,
the decoding can be carried out as follows. If the first bit is 1,
the letter is $a_4$. Otherwise, read the next two bits and treat
them as an integer (in a binary system) denoting the
code of the letter (i.e. 00 corresponds $a_0$, 01 corresponds
$a_1$, etc.) This simple observation can be
generalized and extended for constructing a new coding scheme with the
property that the larger the alphabet size is,
the more speeding-up we get.

In principle, the proposed method can be  applied to the Huffman
code, arithmetic code, and other lossless codes for speeding them
up, but for the sake of simplicity, we will consider the
arithmetic code in the main part of the paper, whereas the Huffman
code  and some others will be mentioned only briefly, because, on
the one hand, the arithmetic code is widely used in practice and,
on the other hand, generalizations are obvious.

The suggested scheme can be applied to sources with unknown
statistics. As we mentioned above, the alphabet letters should be
ordered according to their frequency of occurrences when the
encoding and decoding are carried out. Since the frequencies are
changing after coding of each message letter, the order should be
updated, and the time of such updating should be taken into
account when we estimate the speed of the coding. It turns out
that there exists an algorithm and data structure, which give a
possibility to carry out the updating with few operations per
message letter, and the amount of these operations does not depend
on the alphabet size and/or a probability distribution.

The rest of the paper is organized as follows. The second part
contains estimations of the redundancy caused by the grouping of
letters, and it contains examples for several values of the
redundancy. A fast method of the adaptive arithmetic code for the
grouped alphabet as well as the data structure and algorithm for
easy maintaining the alphabet ordered according to  the frequency
of the occurrences are given in the third and
 the fourth parts. Appendix contains all the proofs.

\section{The redundancy due to grouping. }

First we give some definitions. Let $A = \{ a_1, a_2,\ldots, a_N
\}$ be an alphabet with a probability distribution $\bar{p} = \{
p_1, p_2,\ldots, p_N \}$ where $ p_1 \geq p_2 \geq \ldots \geq
p_N, N \geq 1 $. The distribution can be either known a priori or
it can be estimated from the occurrence counts. In the last case
the order of the probabilities should be updated after encoding
each letter, and it should be taken into account when the speed of
coding is estimated. The simple data structure and algorithm for
maintaining the order of the probabilities  will be described in
the fourth part, whereas here we discuss estimation of the
redundancy.

Let the letters from the alphabet $A$ be grouped as follows : $A_1 =
\{ a_1, a_2,$ $  \ldots, a_{n_1} \},$ $A_2 = \{
a_{n_1+1},a_{n_1+2},\ldots, a_{n_2}  \},\ldots, A_s =  \{
a_{n_{s-1}+1},a_{n_{s-1}+2},\ldots, a_{n_{s}}  \} $ where $n_s =
N, s \geq 1 $. We define the probability distribution $\pi$ and
the vector $\bar{m}= (m_1,$ $ m_2,..., $ $ m_s)$ by
\begin{equation}\label{pi}\pi_i = \sum
_{a_j \in A_i} p_j
\end{equation}
 and $m_i = (n_i - n_{i-1}), n_0 =0, i
= 1, 2, \ldots,s $, correspondingly. In fact,the grouping is
defined by the vector $\bar{m}$. We intend to encode all
letters from one subset $A_i$ by the codewords of equal length.
For this purpose we ascribe equal probabilities to the letters
from $A_i$ by
\begin{equation}\label{code}
 \hat{p}_j = \pi_i / m_i
\end{equation}
 if $a_j \in A_i, i = 1, 2,
\ldots,s.$ Such encoding causes redundancy, defined by
\begin{equation}\label{red}
r(\bar{p}, \bar{m}) = \sum_{i=1}^N p_i \log ( p_i / \hat{p}_i ).
\end{equation}
(Here and below $\log(\:)= \log_2(\:).$)

The suggested method of grouping is based on information about the
order of probabilities (or their estimations). We are
interested in an upper bound for the redundancy (\ref{red})
 defined by
\begin{equation}\label{Red}
\ R( \bar{m})= \sup_{ \bar{p} \in \bar{P }_N} r(\bar{p}, \bar{m})
;  \: \bar{P}_N = \{ p_1, p_2,\ldots, p_N \} : p_1 \geq p_2 \geq
\ldots \geq p_N \}.
\end{equation}
The following theorem gives  the redundancy estimate.

\textbf{Theorem 1.}

{\it The following equality for the redundancy (\ref{Red}) is
valid.
\begin{equation}\label{th}
\ R( \bar{m})= \max_{i=1,...,s} \max_{l=1,...,m_i} l\, \log (m_i
/l)/ (n_i + l),
\end{equation}
where, as before, $\bar{m}= (m_1, m_2,...,m_s), n_i = \sum_{j=1}^i
m_j, i=1, ...,s. $ }

\emph{The proof }is given in Appendix.

The practically interesting question is how to find a grouping
which minimizes the number of groups for a given  upper bound of
the redundancy $\delta$. Theorem 1 can be used as the basis
for such an algorithm. This algorithm
is implemented as a Java program and has been used for preparation
of all examples given below. The program can be found on the
internet and used for practical needs, see

$http://www.ict.nsc.ru/~ryabko/GroupYourAlphabet.html  .$

Let us consider some examples of such grouping  carried
out by the program  mentioned.

First we consider the Huffman code. It should be noted
that in the case of the Huffman code the size of each group should
be a power of 2, whereas it can be any integer
in case of an arithmetic code. This is because the length of
Huffman codewords must be integers whereas this limitation is
absent in arithmetic code.

For example, let the alphabet have 256 letters and let the additional
redundancy (\ref{code}) not exceed 0.08 per letter.
(The choice of these parameters is appropriate, because an alphabet of $2^8 =
256$ symbols is commonly used in compressing computer files, and
the redundancy 0.08  a letter gives 0.01 a bit.) In this case the
following grouping
gives the minimal number of the groups $s$. $$A_1= \{ a_1 \} ,
A_2= \{ a_2 \} , \ldots , A_{12}= \{ a_{12}\}, $$ $$ A_{13}= \{
a_{13}, a_{14}\}, A_{14}= \{ a_{15}, a_{16}\}, \ldots,A_{19}= \{
a_{25}, a_{26}\}, $$ $$A_{20}= \{ a_{27}, a_{28}, a_{29}, a_{30}
\}, \ldots, A_{26}= \{ a_{51}, a_{52}, a_{53}, a_{54} \}, $$  $$
A_{27}= \{ a_{55}, a_{56},\ldots, a_{62} \},\ldots, A_{32}= \{
a_{95},\ldots, a_{102} \}, $$ $$ A_{33}= \{ a_{103},
a_{104},\ldots, a_{118} \},\ldots, A_{39}= \{ a_{199},\ldots,
a_{214} \}, $$ $$ A_{40}= \{ a_{215}, a_{216},\ldots, a_{246} \},
A_{41}= \{ a_{247},\ldots, a_{278} \}. $$ We see that each
of the first 12 subsets contains one letter, each of the subsets
$A_{13}, \ldots, A_{19}$ contains two letters, etc., and the total
number of the subsets $s$ is 41. In reality we could let the last
subset $A_{41}$ contain the letters $\{ a_{247},\ldots, a_{278}
\}$ rather than the letters $ \{ a_{247},\ldots, a_{256} \}$, since each
letter from this subset will be encoded \emph{inside} the subset
by 5- bit words (because $\log 32 = 5$).

Let us proceed with this example in order to show how such a
grouping can be used to simplify the encoding and
 decoding of the Huffman code. If someone knows the
letter probabilities, he can calculate the probability
distribution $\pi$ by (\ref{pi}) and the Huffman code for the new
alphabet $\hat{A} = A_{1}, \ldots, A_{41}$ with the distribution
$\pi$. If we denote a codeword of $A_i$ by $code (A_i)$ and
enumerate all letters in each subset $A_i$ from 0 to $|A_i| -1 $,
then the code of a letter $a_j \in A_i $ can be presented as the
pair of the words $$code (A_i)\: \{number\, of \, a_j \, \in A_i
\},$$ where $ \{number\, of \, a_j \, \in A_i \} $ is the $\log
|A_i|$\,- bit notations of the $a_j$ number (inside $A_i$). For
instance, the letter $a_{103}$ is the first in the 16- letter
subset $A_{33}$ and $a_{246}$ is the last in the 32- letter subset
$A_{40}$. They will be encoded by $code( A_{33})\,0000$ and
$code(A_{40})\,11111$, correspondingly. It is worth noting that
the $code (A_i)\, ,i=1,\ldots, s,$ depends on the probability
distribution whereas the second part of the codewords $\{number\,
of \, a_j \, \in A_i \}$ does not do that. So, in fact, the
Huffman code should be constructed for the 41- letter alphabet
instead of the 256- one, whereas the encoding and decoding inside
the subsets may be implemented with few operations. Of course,
this scheme can be applied to a Shannon code, alphabetical code,
arithmetic code and many others. It is also important that the
decrease of the alphabet size is larger when the alphabet size is
large.

Let us consider one more example of grouping, where the subset
sizes don't need to  be powers of two. Let, as before, the
alphabet have 256 letters and let the additional redundancy
(\ref{code}) not to exceed 0.08 per letter. In this case the
optimal grouping is as follows.  $$ |A_1| = |A_2| =  \ldots ,
|A_{12}| = 1, |A_{13}| = |A_{14}| = \ldots= |A_{16}|= 2, |A_{17}|=
|A_{18}| = 3,
$$
$$|A_{19}|= |A_{20}| = 4, |A_{21}| =5 , |A_{22}| = 6,|A_{23}| = 7,
|A_{24}| = 8, |A_{25}| =9,$$ $$ |A_{26}| = 11,|A_{27}| =
12,|A_{28}| = 14, |A_{29}| = 16, |A_{30}| = 19, $$ $$|A_{31}| =
22, |A_{32}| = 25, |A_{33}| = 29,|A_{34}| = 34,|A_{35}| = 39.$$ We
see that the total number of the subsets (or the size of the
new alphabet) is less than in the previous example (35 instead of
41), because in the first example the subset sizes should be
powers of two, whereas there is no such limitation in the
second case. So, if someone can accept the additional redundancy
0.01 per bit, he can use the new alphabet $ \hat{A} = \{ A_{1},
\ldots, A_{35} \} $ instead of 256- letter alphabet and implement
the arithmetic coding in the same manner as it was described for
the Huffman code. (The exact description of the method will be
given in the next part). We will not consider the new examples in
details, but note again that the decrease in the number of the
letters is more, when the alphabet size is larger. Thus, if the
alphabet size is $2^{16}$ and the redundancy upper bound is 0.16
(0.01 per bit), the number of groups $s$ is 39, and if the size is
$2^{20}$ then $s= 40 $ whereas the redundancy per bit is the same.
(Such calculations can be easily carried out by the above
mentioned program).

The required grouping for decreasing the
alphabet size is based on the simple theorem 2, for which
 we need to give some definitions standard in source
coding.

Let $\gamma$ be a certain method of source coding which can be
applied to letters from a certain alphabet $A$. If $p$ is a
probability distribution on $A$, then the redundancy of $\gamma$
and its upper bound are defined by
\begin{equation}\label{red1}
\rho(\gamma, p) = \sum_{a \in A} p(a)( |\gamma (a) |+ \log p(a)),
\quad \hat{\rho}(\gamma ) = sup_{p} \:\rho(\gamma, p),
\end{equation}
where the supremum is taken over all distributions  $p$, $|\gamma
(a) |$ and $p(a)$ are the length of the code word and the
probability of $a \in A$, correspondingly. For example,
$\hat{\rho} $ equals 1 for the Huffman and the Shannon codes
whereas for the arithmetic code $\hat{\rho}$ can be done as small
as it is required by choosing some parameters, see, for ex.,
\cite{Ryabko-Fionov}.

The following theorem gives a formal justification for applying
the above described grouping for source coding.

\textbf{Theorem 2.} {\it Let  the redundancy of a certain code
$\gamma$ be not  more than some $\Delta$ for all probability
distributions. Then, if the alphabet is divided into subsets $A_i,
i=1,\ldots, s ,$ in such a way that the additional redundancy
(\ref{red}) equals $\delta$, and the code $\gamma$ is applied to
the probability distribution $\hat{p}$ defined by (\ref{code}),
then the total redundancy of this new code $\gamma_{gr}$ is upper
bounded by $\Delta+\delta$.}

Theorem 1 gives a simple algorithm for finding the grouping
which gives the minimal number of the groups $s$ when the upper
bound for the admissible redundancy  (\ref{Red}) is given. On the
other hand, the simple asymptotic estimate of the number of
such groups and the group sizes can be interesting when the
number of the alphabet letters is large. The following theorem can
be used for this purpose.

\textbf{Theorem 3.}

{\it Let $\delta > 0 $ be an admissible redundancy (\ref{Red}) of
a grouping.

i) If
\begin{equation}\label{co1}
\quad m_i \,\leq \,\lfloor \,\delta\, n_{i-1}\,\, e \,/ (\log e -
\delta \,e)\,\rfloor,
\end{equation}
then the redundancy of the grouping $(m_1, m_2, \ldots )$ does not
exceed $\delta$, where $n_i = \sum_{j=1}^i\, m_j, \;$ $e\approx
2.718... .$).

ii) the minimal number of groups $\:s\:$ as a function of the
redundancy $\delta$ is upper bounded by
\begin{equation}\label{co}
c \log N / \delta + c_1,
\end{equation}
where  $c$ and $c_1$ are constants and $N$ is the alphabet
size,$\:N \rightarrow \infty.$ }

\emph{The proof} is given in Appendix.

\textbf{Comment 1.} {\it The first statement of the theorem 3
gives
 construction of the $\delta-$ redundant grouping $(m_1,
m_2, ...)$ for an infinite alphabet, because $m_i$ in (\ref{co1})
depends only  on previous $m_1, m_2, \ldots, m_{i-1}$.}

\textbf{Comment 2.} {\it Theorem 3 is valid for grouping where the
subset sizes $(m_1, m_2, \ldots )$ should be powers of 2. }

\section{The arithmetic code for grouped alphabets. }

Arithmetic coding was introduced by Rissanen \cite{Riss76} in 1976
and now it is  one of the most popular methods of source coding,
see, e.g., \cite{Moffat94}, \cite{Ryabko-Fionov}. The advantage of
arithmetic coding over other coding techniques is that it achieves
arbitrarily small coding redundancy per source symbol at less
computational effort than any other method.

We give first a brief description of an arithmetic code by paying
attention to features which determine the speed of encoding and
decoding. As before, consider a memoryless source generating
letters from the alphabet $A= \{ a_1, ..., a_{N} \}$ with unknown
probabilities. Let the source generate a message $x_1\ldots
x_{t-1}x_t\ldots $, $x_i\in A$ for all $i$, and let $ \nu^t(a)$
denote the occurrence count of letter $a$  in the word $x_1\ldots
x_{t-1}x_t $. After
 first $t$ letters $x_1,\ldots, x_{t-1},x_t$ have been processed
the following letter $ x_{t+1}$ needs to be encoded. In the most
popular version of the arithmetic code the  current estimated
probability distribution is taken as
\begin{equation}\label{piti}
 p^t(a)= (\nu^t(a)+c)/(t+Nc) , a \in A ,
\end{equation}
where $c$ is a constant (as a rule $c$ is 1 or 1/2). Let $x_{t+1}=
a_i$, and let the interval $[\alpha, \beta )$ represent the word
$x_1 \ldots x_{t-1} x_t $. Then the word $x_1 \ldots x_{t-1} x_t
x_{t+1}$, $x_{t+1}= a_i $ will be encoded by the interval
\begin{equation}\label{int}
[\alpha + ( \beta - \alpha)\: q^t_i,\quad  \alpha + ( \beta -
\alpha)\: q^t_{i+1}\: )\,,
\end{equation}
 where
\begin{equation}\label{qu}
q^t_i = \sum _{j=1}^{i-1} p^t(a_j).
\end{equation}
When the size of the alphabet $N$ is large, the calculation of
$q^t_i$ is the most time consuming part in the encoding process.
As it was mentioned in the introduction, there are fast algorithms
for calculation of $q^t_i$ in
\begin{equation}\label{time}
T= c_1 \log N + c_2,
\end{equation}
operations under $ (\log N + \tau)$- bit words, where $\tau$ is
the constant determining the redundancy of the arithmetic code.
(As a rule, this length is in proportional to the length of the
computer word: 16 bits, 32 bits, etc.)

We describe a new algorithm for the alphabet whose letters are
divided into subsets $ A_1^t,\ldots, A_s^t, $ and the same
probability is ascribed to all letters in the subset. Such a
separation of the alphabet $A$ can depend on $t$ which is why the
notation $A_i^t$ is used. But, on the other hand, the number of
the letters in each subset $A_i^t$ will not depend on $t$ which is
why it is denoted as $|A_i^t| = m_i$.

In principle, the scheme for the arithmetic coding is the same as
in the above considered case of the Huffman code: the
codeword of the letter $ x_{t+1}= a_i $ consists of two parts,
where the first part encodes the set $A^t_k$ that contains $a_i$,
and the second part encodes the ordinal of the element $a_i$ in the
set $A^t_k$. It turns out that it is easy to encode and decode
letters in the sets $A^t_k$, and the time
consuming operations should be used to encode the sets $A^t_k$, only.

We proceed with the formal description of the algorithm. Since the
probabilities of the letters in $A$ can depend on $t$ we define in
analogy with (\ref{pi}),(\ref{code})
\begin{equation}\label{PQ1}
\pi_i^t = \sum _{a_j \in A_i} p_j,\quad \hat{p}_i^{\,t} = \pi_i^t
/ m_i
\end{equation}
and let
\begin{equation}\label{PQ}
 Q^t_i=  \sum _{j=1}^{i-1} \pi_j^t\:.
\end{equation}

The arithmetic encoding and decoding are implemented for the
probability distribution (\ref{PQ1}), where the probability
$\hat{p}_i^{\,t}$ is ascribed to all letters from the subset
$A_i$. More precisely, assume that the letters in each $A^t_k$ are
enumerated from 1 to $m_i$, and that the encoder and the decoder
know this enumeration. Let, as before, $ x_{t+1}= a_i $, and let
$a_i$ belong to $A^t_k$ for some $k$. Then the coding interval for
the word $x_1\ldots x_{t-1}x_t x_{t+1}$ is calculated as follows
\begin{equation}\label{newint}
[\alpha + ( \beta - \alpha)( Q^t_k + (\delta (a_i)-1)\,
\hat{p}_i^{\,t}\,)\, ,\quad
 \alpha + ( \beta - \alpha) ( Q^t_k + \delta (a_i)\,\hat{p}_i^{\,t})\; ),
\end{equation}
where $ \delta(a_i)$ is the ordinal of $a_i$ in the subset
$A^t_k$. It can be easily seen that this definition is equivalent
with (\ref{int}), where the probability of each letter from $A_i$
equals $  \hat{p}_i^{\,t}$.
 Indeed, let us order the letters of $A$
according to their count of
 occurrence in the word $x_1\ldots x_{t-1}x_t, $ and let the letters
 in  $A^t_k,k=1,2,...,s\, ,$ be ordered according to the
 enumeration mentioned above. We then immediately obtain
 (\ref{newint}) from  (\ref{int}) and (\ref{PQ1}). The additional redundancy which
 is caused by the replacement of the distribution (\ref{piti}) by
 $ \hat{p}_i^{\,t}$  can be estimated using (\ref{red}) and the theorems 1-3,
which is why
 we may
 concentrate our attention on the  encoding and decoding speed
 and the storage space needed.

First we  compare the time needed for the calculation in
(\ref{int}) and (\ref{newint}). If we ignore the expressions
$(\delta (a_i)-1) \hat{p}_i^{\,t}$ and $ \delta (a_i)
\hat{p}_i^{\,t}$ for a while, we see that (\ref{newint}) can be
considered as the arithmetic encoding of the new alphabet $ \{
A^t_1$, $A^t_2,...,$ $A^t_s \} $. Therefore, the number of
operations for encoding by (\ref{newint}) is the same as the time
of arithmetic coding for the $s$ letter alphabet, which by
(\ref{time}) equals $c_1 \log s + c_2 $. The expressions $(\delta
(a_i)-1)\hat{p}_i^{\:t}$ and $ \delta (a_i) \hat{p}_i^{\:t}$
require two multiplications, and two additions are needed to
obtain bounds of the interval in (\ref{newint}). Hence, the number
of operations for encoding ($T$) by (\ref{newint}) is given by
\begin{equation}\label{newtime}
T= c_1^* \log s + c_2^* ,
\end{equation}
where $c_1^*, c_2^*$ are constants and all operations are carried
out under the word of the length  $ (\log N + \tau)$- bit as it
was required for the usual arithmetic code. In case $s$ is much
less than $N$, the time of encoding in the new method is less than
the time of the usual arithmetic code, see (\ref{newtime}) and
(\ref{time}).

We describe shortly decoding with the new method. Suppose that the
letters $x_1 \ldots x_{t-1} x_t $ have been decoded and the letter
$x_{t+1}$ is to be decoded.
 There are two steps required:
first, the algorithm finds  the set $A^t_k$ with the usual
arithmetic code that contains the (unknown) letter $a_i$. The
ordinal of the letter $a_i$ is calculated as follows:
\begin{equation}\label{decode}
\delta ( ) = \lfloor(code (x_{t+1}...) - Q^t_j )/
\hat{p}_i^{\,t}\rfloor,
\end{equation}
where $ code (x_{t+1}...)$ is the number that encodes the word
$x_{t+1}x_{t+2}...$. It can be seen that (\ref{decode}) is the
inverse of (\ref{newint}). In order to calculate (\ref{decode})
the decoder should carry out one division and one subtraction.
That is why the  total number of decoding operations is  given by
the same formula as for the encoding, see (\ref{newtime}).

It is worth noting that multiplications and divisions in
(\ref{newint}) and  (\ref{decode}) could be carried out faster if
the subset sizes are powers of two. But, on the other hand, in
this case the number of the subsets is larger, that is why both
version could be useful.

We did not estimate yet the time needed for maintaining the order
of letters from $A$ according to their frequencies (\ref{piti}).
The point is that the order should be updated by the encoder and
the decoder after encoding and decoding each letter $x_t$. It
turns out that it is possible  to update the order using a fixed
number of operations. Such a method is described in the next
section. Besides, we should take into account that, when $x_t$ is
encoded (or decoded), one frequency (\ref{piti}) should be changed
and at most two $\pi_i$ (\ref{PQ1}) must be recalculated. It is
easy to see that all these transformations can be done with no
more than two additions and two subtractions. Therefore, the
total number of operations for encoding and decoding is given by
(\ref{newtime}) with the new constant $c_2^*$.

So we can see that
%
if the arithmetic code can be applied to an $N \:- $ letter source, so
that the number of operations (under words of a certain length) of
coding is $$ T= c_1 \log N + c_2,$$ then there exists an algorithm
of coding, which can be applied to the grouped alphabet $
A_1^t,\ldots, A_s^t $ in such a way that, first, at each moment
$t$ the letters are ordered by decreasing
frequencies and, second, the number of coding operations is  $$
T= c_1 \log s + c_2^* $$ with words of the same length, where $
c_1, c_2, c_2^* $ are constants.

\section{ A fast algorithm for keeping the alphabet letters ordered.}
 In this section we describe a data structure and an algorithm, which allow
one to carry out all the operations for maintaining the alphabet letters
ordered by their frequencies, in such a way that the
number of such operations is constant, independently of the
probability distribution, the size of the alphabet, and other
characteristics.

The  data structure suggested is based on five  arrays $Fr [1 :
N], Sorted$ $ Alphabet [1:N],$ $ Inverse Sort[1:N], SetBegin [0:
MAX ], SetEnd [0: MAX ]$, where, as before, $N$ is the size of the
alphabet, $\Lambda^t_k$ is the set of the letters from $A$, which
frequency of the occurrence equals $k$ at the moment $t$ and $MAX$
is an upper bound for the maximal count of occurrence (For
example, if the code uses the sliding window to adapt to the
source, $MAX$ is upper bounded by the length of the window). At
each moment $t$ the array $Fr$ contains information about
frequencies of occurrence of the letters from $A$ in the word $
x_1 \ldots x_{t-1} x_t$  such that $Fr[i]= \nu^t(a_i)$. The array
$SortedAlphabet [1:N]$ consists of letters from $A$ ordered by the
frequency of occurrence. More precisely, the following property is
satisfied: if $i \leq j$ and $ Sorted Alphabet [i]= b $ and $
Sorted Alphabet [j]= c$, then $ \nu^t(b) \leq \nu^t(c)$. In
particular, it means that all letters from a subset $\Lambda^t_k,
k=0,1,...$, are situated in succession in $Sorted Alphabet [1:N]$
and forming a string. $SetBegin [k ]$ and $SetEnd [k]$ contain
information about the beginning and the end of such a string.
 At last, by definition,$ Inverse Sort[i]$
contains  an integer $j$ such that $Sorted Alphabet [j]= a_i$.

Let us consider a small example. Let $N = 4 $, $t = 4 $ and  the
frequencies
 $\nu^t(a_1)=0, \nu^t(a_2)=1, \nu^t(a_3)=2 $ and $ \nu^t(a_4)=1 $.
Then, $Fr= $ $ [0,1,2,1],$ $ Sorted Alphabet $ $ =
[a_1,a_4,a_2,a_3],$ $ Inverse Sort =[1,3,4,2]$,$Set $ $ Begin$ $
=[1,2,4]$, $ SetEnd $ $ =[1,3,4] $ is one possible configuration
of the contents of the relevant arrays.

Consider next updating the information in the arrays, which should
be done by the encoder (and decoder) after encoding (and decoding)
of each letter, in such a way that only a constant number of
operations is needed. Suppose we encode the letter $a_4$ and
increment its occurrence count. The arrays should be changed as
follows : the processed letter ($a_4$) should be exchanged with
the last letter from $ \Lambda^t_k $ ($ \Lambda^t_1 $ in our case)
and the relevant modifications should be done in $ SortedAlphabet$
and $InverseSort $. Then the letter processed should be included
in the set $ \Lambda^t_{k+1} $ and excluded from the set $
\Lambda^t_k $. In fact, it is enough to change two elements in
$SetBegin$ and $ SetEnd $, namely, $SetBegin [k+1]=
SetBegin[k+1]-1 $ and $ SetEnd[k]= SetEnd [k]- 1 $.  (In our
example, $a_4$ should be moved from $ \Lambda^t_1 $ into $
\Lambda^t_2 $. When we carry out these calculations the result is
$Fr= [0,1,2,2],$ $ SortedAlphabet = [a_1,a_2,a_4,a_3],$ $
InverseSort=[1,2,4,3], $ $ SetBegin =[1,2,3] $ and $ SetEnd
=[1,2,4] $.)

We have considered the case when the occurrence count should be
incremented. Decrementing, which is used in certain schemes of the
adaptive arithmetic code, can be carried out in a similar manner.

\section{Appendix. }

\textbf{The proof of Theorem 1.} It is easy to see that the set
$\bar{P}_N$ of all  distributions  which are ordered according to
the probability decreasing is convex. Indeed, each $ \bar{p} = \{
p_1, p_2,\ldots, p_N \} \in \bar{P}_N$ may be presented as a
 linear combination of vectors from the set
 \begin{equation}\label{q}
Q_N = \{q_1 = (1,0,\ldots,0), q_2= (1/2,1/2,0,\ldots,0),\ldots,
q_N = ( 1/N,   \ldots, 1/N)
\end{equation}
 as follows:
$$ \bar{p} = \sum_{i=1}^N (p_i - p_{i+1 } ) q_i ;  $$
where $p_{N+1}= 0 .$

On the other hand, the redundancy (\ref{red}) is a convex function,
because the direct calculation shows that its second partial
derivatives are nonnegative. Indeed, the redundancy (\ref{red})
can be represented as follows. $$ r(\bar{p}, \bar{m}) = \sum_{i=1
}^N p_i \log ( p_i ) \: - \,\sum_{j=1}^s \pi_j (\log \pi_j - \log
m_j) = $$

$$ \sum_{i=2 }^N p_i \log ( p_i )  \:- \,
\sum_{j=2}^s \pi_j (\log \pi_j - \log m_j)\, +$$
$$(1-\sum_{k=2}^N p_k ) \log (1-\sum_{k=2}^N p_k )\,
-\,(1-\sum_{l=2}^s \pi_l )
( \log (1-\sum_{l=2}^s \pi_l ) - \log m_1). $$ If $a_i$ is a
certain letter from $A$ and $j$ is such a subset that $a_i \in A_j
$ then, the direct calculation shows that
$$ \partial r / \partial p_i = \log_2 e \,(\,\ln p_i - \ln \pi_j-
\ln (1 - \sum_{k=2}^N p_k) + \ln (1 - \sum_{l=2}^s \pi_l)\,) +
constant ,
$$

$$\partial^2 r /
\partial^2 p_i = \log_2 e \,((- 1/\pi_i + 1/ p_j) +
(- 1/\pi_1 + 1/ p_1)) .$$ The last value is nonnegative, because,
by definition, $\pi_i = \sum _{k= n_i }^{n_{i+1}-1}p_k$ and $p_j$
is one of the summands as well as $p_1$ is one of the summands of
$\pi_1$.

Thus, the redundancy is a convex function defined on a
convex set, and its  extreme points are $Q_N$ from (\ref{q}). So
$$sup_{ \bar{p} \in \bar{P }_N} r(\bar{p}, \bar{m}) = \max_{ q
\,\in \;Q_N} r(q, \bar{m})  .$$ Each  $q \in Q_N$ can be presented
as a vector $ q= (1/(n_i + l), \ldots, 1/(n_i + l), 0, \ldots, 0 )
$ where $ 1 \leq l \leq m_{i+1} , i=0, \ldots, s-1.$ This
representation, the last equality, the definitions (\ref{q}) ,
(\ref{red}) and (\ref{Red}) give (\ref{th}).

\textbf{Proof of the theorem 2.} Obviously,
\begin{equation}\label{obv}
\sum_{a \in A} p(a)( |\gamma_{gr} (a) |+ \log p(a)) =$$ $$ \sum_{a
\in A} p(a)( |\gamma_{gr}(a) |+ \log \hat{p}(a)) + \sum_{a \in A}
p(a)(
 \log (p(a)/ \hat{p}(a)).
\end{equation}
and, from (\ref{pi}),(\ref{code}) we obtain $$ \sum_{a \in A}
p(a)( |\gamma_{gr}(a) |+ \log \hat{p}(a))= \sum_{i=1}^s (
|\gamma_{gr}(a) |+ \log \hat{p}(a)) \sum_{a \in A_i} p(a) = $$ $$
\sum_{i=1}^s ( |\gamma_{gr}(a) |+ \log \hat{p}(a)) \sum_{a \in
A_i} \hat{p}(a) = \sum_{a \in A} \hat{p}(a)( |\gamma_{gr}(a) |+
\log \hat{p}(a)). $$ This equality and (\ref{obv}) gives
$$
\sum_{a \in A} p(a)( |\gamma_{gr} (a) |+ \log p(a)) =$$ $$ \sum_{a
\in A} \hat{p}(a)( |\gamma_{gr}(a) |+ \log \hat{p}(a)) + \sum_{a
\in A} p(a)(
 \log (p(a)/ \hat{p}(a)).$$ From this equality, the statement of the theorem and
the definitions (\ref{red}) and (\ref{red1}) we obtain
$$
\sum_{a \in A} p(a)( |\gamma_{gr} (a) |+ \log p(a)) \leq \Delta +
\delta.
$$
Theorem 2 is proved.

\textbf{The proof of the theorem 3.} The proof is based on the
theorem 1. From (\ref{th}) we obtain the following obvious
inequality
 \begin{equation}\label{cr}
R( \bar{m})\leq \max_{i=1,...,s} \max_{l=1,...,m_i} l\, \log (m_i
/l)/ n_i .
\end{equation} Direct calculation shows that
$$
\partial (\log (m_i /l)/n_i)/\partial l = \log_2 e \,(\ln
(m_i/l) - 1 )/n_i ,$$
$$ \partial^2(\log (m_i /l)/n_i)/\partial l^2 = - \log_2e/(l \,n_i)
<0
$$ and, consequently, the maximum of the function
$\log (m_i /l)/n_i$ is equal to $ m_i\log e / (e \,n_i) ,$ when
$l= m_i/e $. So,
$$ \max_{l=1,...,m_i} l\, \log (m_i/l)/ n_i \leq m_i\log e / (e\,
n_i) $$ and from (\ref{cr}) we obtain
\begin{equation}\label{cr1}
R( \bar{m})\leq \max_{i=1,...,s} m_i\log e / (e \,n_i).
\end{equation}
That is why, if
\begin{equation}\label{cr2}
m_i \leq \delta \,e \,n_i/ \log e
\end{equation}
 then $R( \bar{m})\leq \delta $. By
definition ( see the statement of the theorem ) , $n_i = n_{i-1} +
m_i$ and we obtain from (\ref{cr2}) the first claim of the
theorem. Taking into account that $n_{s-1} < N \leq n_s $ and
(\ref{cr1}), (\ref{cr2}) we can see that, if
$$ N = \acute{c}_1 (1+\delta e/ \log e)^s + \acute{c}_2, $$  then
$R( \bar{m})\leq \delta ,$ where $ \acute{c}_1$ and $\acute{c}_2$
are constants and $N \rightarrow \infty .$ Taking the logarithm
and applying the well known estimation $\ln (1+\varepsilon )
\approx \varepsilon$ when $ \varepsilon \approx 0, $ we obtain
(\ref{co}). The theorem is proved.





\newpage

\noindent{\bf Authors:}

\noindent B.Ya. Ryabko\\
 Professor.\\
 Siberian State University of Telecommunication and Computer Science\\
Kirov Street, 86\\
630102 Novosibirsk, Russia.
 \vskip .05in
\noindent e-mail: \verb"ryabko@neic.nsk.su" \\
URL: \verb"www.ict.nsc.ru/~ryabko"

\vskip .1in

\noindent J. Astola \\
Professor.\\
Tampere University of Technology\\
P.O.B. 553, FIN- 33101 Tampere,\\ Finland.
 \vskip .05in \noindent
e-mail: \verb"jta@cs.tut.fi"

\vskip .1in

\noindent K. Eguiazarian \\
Professor.\\
Tampere University of Technology\\
P.O.B. 553, FIN- 33101 Tampere, \\ Finland.
 \vskip .05in
\noindent e-mail: \verb"karen@cs.tut.fi"

\vskip .1in
\noindent{\bf  Address for Correspondence:}\\
\noindent  prof. B. Ryabko \\
  Siberian State University of Telecommunication and
Computer
Science\\
Kirov Street, 86\\
630102 Novosibirsk, Russia.\\
\noindent e-mail: \verb"ryabko@neic.nsk.su" \\


\newpage
\begin{thebibliography}{5}

\bibitem{Aho}
A.V.Aho,J.E. Hopcroft, J.D.Ulman.{ \it The desighn and analysis of
computer algorithms }, Reading, MA: Addison- Wesley, 1976.

\bibitem{Fenwick}
P. Fenwick, ``A new data structure for cumulative probability
tables,'' {\it Software -- Practice and Experience,} vol. 24, no.
3, pp. 327--336, March 1994. Errata published in vol. 24, no. 7,
p. 667, July 1994.

\bibitem{Jo}
D. W. Jones", "Application of splay trees to data compression",
        {\it Communications of the ACM}, v 31, n. 8,1988,
        pp. "996-1007",


\bibitem{Ki1}
Kieffer, J.C.; Yang, E.H. Grammar-based codes: a new class of
universal lossless source codes. {\it IEEE Trans. Inform. Theory},
v.46 (2000), no. 3, 737--754.

\bibitem{Ki2}
Kieffer, J.C.; Yang, E.H.; Nelson, G.J.; Cosman, P. Universal
lossless compression via multilevel pattern matching.{\it IEEE
Trans. Inform. Theory,} v.46 (2000), no. 4, 1227--1245.

\bibitem{Moffat90}
A. Moffat, Linear time adaptive arithmetic coding",{\it  IEEE
Transactions on Information Theory }
 1990,    v.36, no. 2, pp.401-406.



\bibitem{Moffat99}
A. Moffat, An improved data structure for cumulative probability
tables, 1999,{\it Software -- Practice and Experience},
        v.29,
        no. 7,
        pp.647-659.


\bibitem{Moffat94}
A.Moffat,R.Neal,I.Witten. "Arithmetic Coding Revisited",  {\it ACM
Transactions on Information Systems,} 16(3):256-294, July 1998.

\bibitem{T-M}
Moffat, A.; Turpin, A. On the implementation of minimum redundancy
prefix codes, {\it  IEEE Transactions on Communications,} v.45,
no. 10, pp. 1200 - 1207, 1997.

\bibitem{M-T1}
A.Moffat,A.Turpin,  Efficient Construction of Minimum-Redundancy
Codes for Large Alphabets. {\it IEEE Trans. Inform. Theory,} vol.
IT-44, no. 4, pp. 1650--1657, July 1998.


\bibitem{Riss76}
J.Rissanen, ``Generalized Kraft inequality and arithmetic
coding,'' {\it IBM J. Res. Dev.,} vol. 20, pp. 198--203, May 1976.



\bibitem{RyabkoDAN}
Ryabko, B. Ya. A fast sequential code. {\it  Dokl. Akad. Nauk SSSR
} v.306 (1989), no. 3, pp.548--552 (Russian); translation in {\it
Soviet Math. Dokl.}, v. 39 (1989), no. 3, pp. 533--537.



\bibitem{Ryabko}
B.Ryabko, A fast on-line adaptive code, {\it IEEE Trans. Inform.
Theory,} vol. IT-38, no. 4, pp. 1400--1404, 1992.



\bibitem{Ryabko-Fionov}
B.Ryabko, A.Fionov, ''Fast and Space-Efficient Adaptive Arithmetic
Coding'',{ \it in :Cryptography and Coding, 7th IMA International
Conference, Cirencester, UK, December 1999. Proceedings }, LNCS
1746, pp. 270 -279.

\bibitem{R-Ri}
B.Ryabko, J.Rissanen. " Fast Adaptive Arithmetic Code for Large
Alphabet Sources
 with Asymmetrical Distributions", { \it IEEE Communications Letters,}
2002, (accepted for publication).

See also B. Ryabko, J. Rissanen. Fast Adaptive Arithmetic Code for
Large Alphabet Sources with Asymmetrical Distributions . { \it
Proceedings of the IEEE International Symposium on Information
Theory, 2002, Lausanne, Switzelend, } p.319


\bibitem{M-T}
Turpin, A.; Moffat, A., ''On-line adaptive canonical prefix coding
with bounded compression loss'',{\it IEEE Trans. Inform. Theory,}
vol. IT-47, no. 1, pp.88- 98, 2001.


\end{thebibliography}
\end{document}